\documentclass[a4paper,11pt]{article}
\usepackage{ILD}
\usepackage[symbol]{footmisc}
\usepackage{feynmf}
\usepackage{array,multirow,graphicx}
\usepackage{graphicx}
\usepackage{cleveref}
\usepackage{tikz}
\usepackage[compat=1.1.0]{tikz-feynman}
\usetikzlibrary{arrows,shapes,positioning}
\usetikzlibrary{decorations.markings}
\usetikzlibrary{decorations.pathmorphing}
\usetikzlibrary{decorations.pathreplacing}
\usetikzlibrary{patterns}
\usetikzlibrary{plotmarks}
\usetikzlibrary{shadows}
\usepackage{contour}
\usepackage[style=numeric-comp,sorting=none]{biblatex}

\usepackage{subcaption}
\usepackage{wrapfig}

\title{Kinematic Fitting for Particle Flow Detectors at Future Higgs Factories}

\ildproc{phys}{2021}{016}

\date{\today}

\addauthor{Yasser Radkhorrami}{\institute{1}\institute{2}}
\addauthor{Jenny List}{\institute{1}}

\addinstitute{1}{Deutsches Elektronen-Synchrotron (DESY), Hamburg}
\addinstitute{2}{Universit\"at Hamburg, Hamburg}

\abstract{In many analyses in Higgs, top and electroweak physics, the kinematic reconstruction of the final state is improved by constrained fits. This is a particularly powerful tool at $e^{+}e^{-}$ colliders, where the initial state four-momentum is known and can be employed to constrain the final state. A crucial ingredient to kinematic fitting is an accurate estimate of the measurement uncertainties, in particular for composed objects like jets. This contribution will show how the particle flow concept, which is a design-driver for most detectors proposed for future Higgs factories, can --- in addition to an excellent jet energy measurement --- provide detailed estimates of the covariance matrices for each individual particle flow object (PFO) and each individual jet. Combined with information about leptons and secondary vertices in the jets, the kinematic fit enables to correct $b$- and $c$-jets for missing momentum from neutrinos from semi-leptonic heavy quark decays. The impact on the reconstruction of invariant di-jet masses and the resulting improvement in $ZH$ vs $ZZ$ separation will be presented, using the full simulation of the International Large Detector (ILD), as an example of highly-granular ParticleFlow optimized detector concept.}

\titlecomment{Based on the talk "Kinematic Fitting for ParticleFlow Detectors at Future Higgs Factories" presented at Particles and Nuclei International Conference (PANIC2021) on behalf of the ILD concept group, Laboratory for Instrumentation and Experimental Particle Physics (LIP) and Faculty of Sciences of the University of Lisbon (FCUL), Lisbon, Portugal, 5-10 September 2021, 398}

\graphicspath{ {./logos/}{./figures/} }

\begin{document}

\titlepage

\section{Introduction}\label{SEC:Intro}
The decays of the Higgs boson into heavy $b$ and $c$ quarks play an essential role in the Higgs physics studies. The presence of the semi-leptonic decays (SLD) in the $b$ and $c$ jets degrades the reconstruction of these jets \cite[fig.~8.3d]{theildcollaboration2020international}. The energy resolution of the jets can be recovered beyond the detector resolution by using a kinematic fit as a mathematical approach \cite{List:88030}. The measurement uncertainties are one of the most important inputs to the kinematic fit. In the ILD, individual particles are reconstructed with an unprecedented knowledge of jet-level uncertainties \cite{theildcollaboration2020international}.\\
In the last study of the Higgs self-coupling measurement at the 500 GeV ILC \cite{Duerig:310520}, the invariant di-jet mass in $e^{+}e^{-}\rightarrow ZHH$ events with $H\rightarrow b\bar{b}$ has been improved by a very simple correction to the semi-leptonic decays and a simple parameterization of the jet energy error in the kinematic fit.
\begin{figure}[htbp] \centering
	\begin{subfigure}[t]{0.49\textwidth}
		\includegraphics[width=0.99\textwidth]{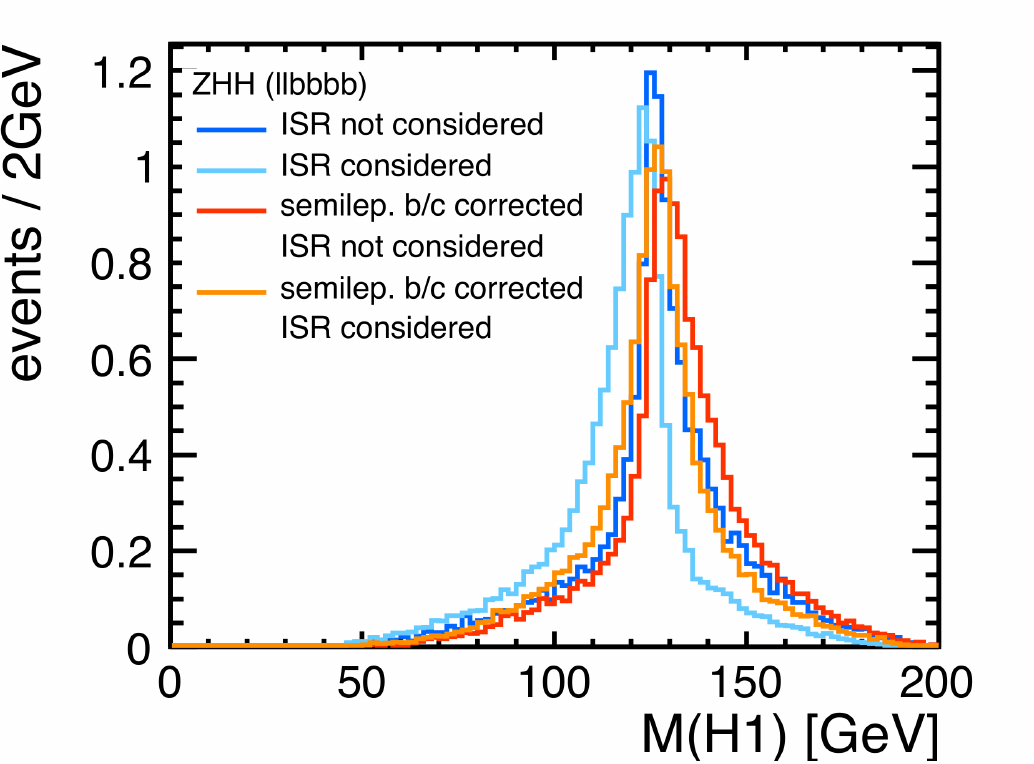}
		\caption{}
		\label{fig:ZHH_Higgsmass}
	\end{subfigure}
	\begin{subfigure}[t]{0.49\textwidth}
		\includegraphics[width=0.99\textwidth]{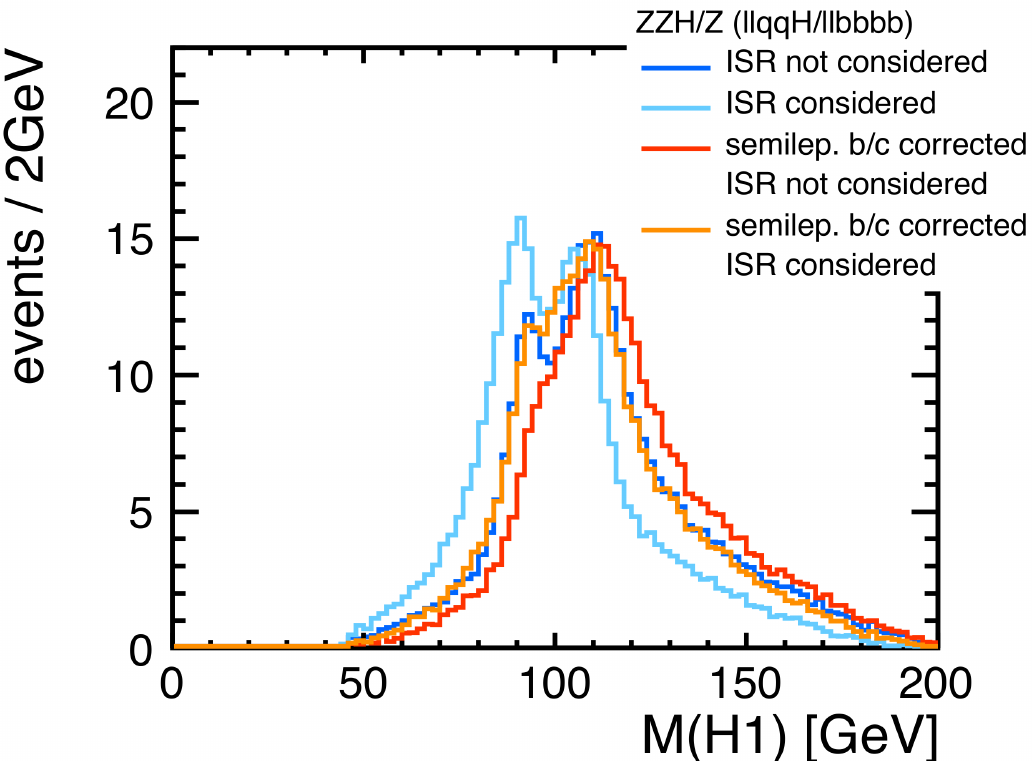}
		\caption{}
		\label{fig:ZZHZ_HiggsmassBG}
	\end{subfigure}
	\caption{Higgs mass reconstruction in the presence of ISR photon and semi-leptonic decays: (a) The signal peak gets sharper by considering ISR photon and correction of semi-leptonic decays. (b) The background peak is pulled towards the signal.}
	\label{fig:ZZHZ_Higgsmass}
\end{figure}
Although the reconstruction of the Higgs mass is improved by this simple correction (fig.~\ref{fig:ZHH_Higgsmass}), the result is not satisfactory in the presence of $ZZH/Z$ background (fig.~\ref{fig:ZZHZ_HiggsmassBG}). In order to avoid background being pulled towards the signal by too much flexibility in the fit, a proper neutrino correction and a better jet error parameterization are needed. The new jet-error parameterization has been evaluated using $ZH/ZZ\rightarrow\mu\bar{\mu}b\bar{b}$ samples in presence of ISR photon and $\gamma\gamma\rightarrow$ low $p_{T}$ hadrons at $\sqrt{s}$=250 GeV \cite{radkhorrami2021conceptual,radkhorrami2021kinematic}.
\section{Kinematic fitting}\label{SEC:KinFit}
\begin{figure}[htbp] \centering
	\begin{subfigure}{0.49\textwidth}
		\includegraphics[width=0.9\textwidth]{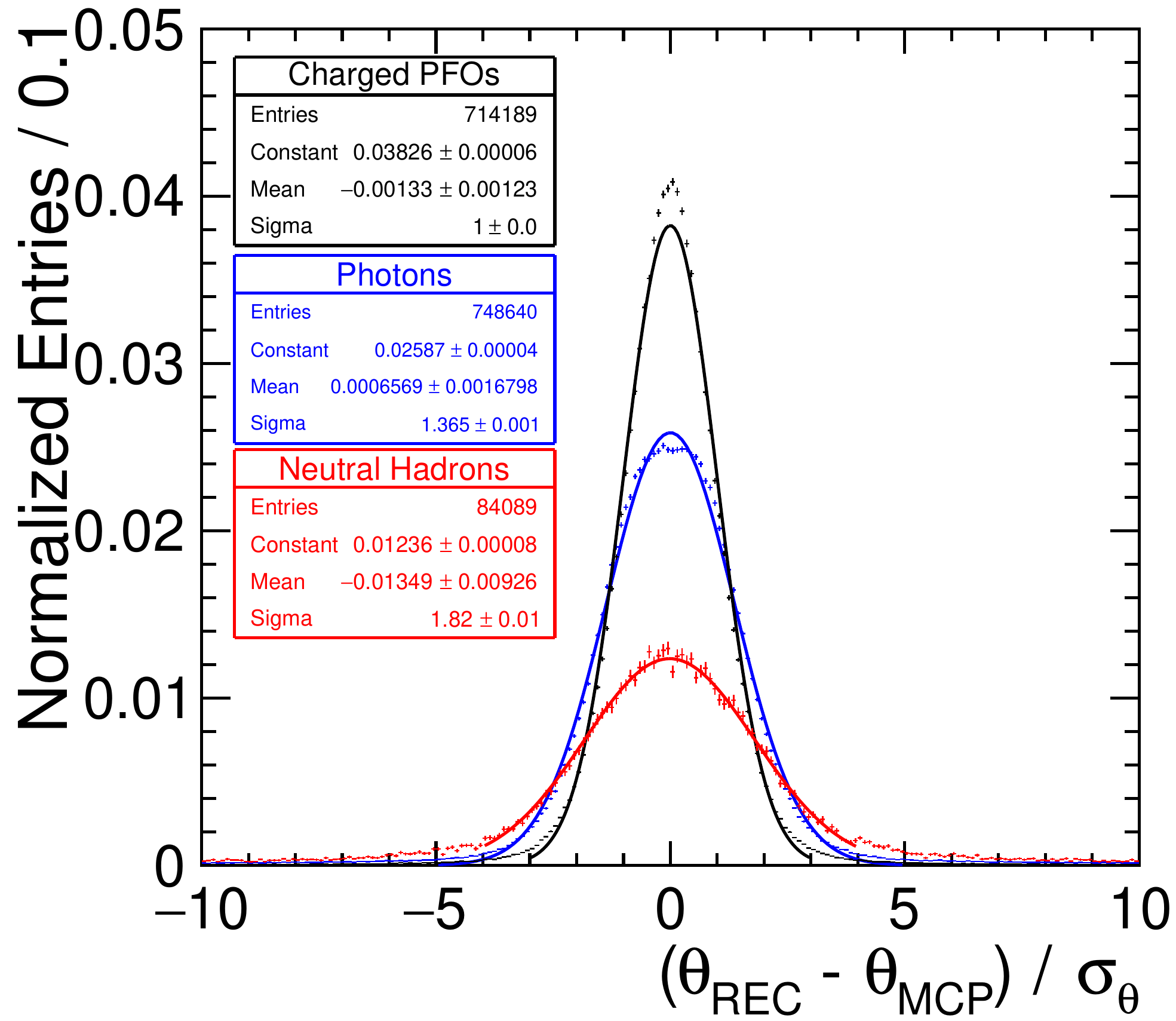}
		\caption{}
		\label{fig:sigmaTheta_pfo}
	\end{subfigure}
	\begin{subfigure}{0.49\textwidth}
		\includegraphics[width=0.9\textwidth]{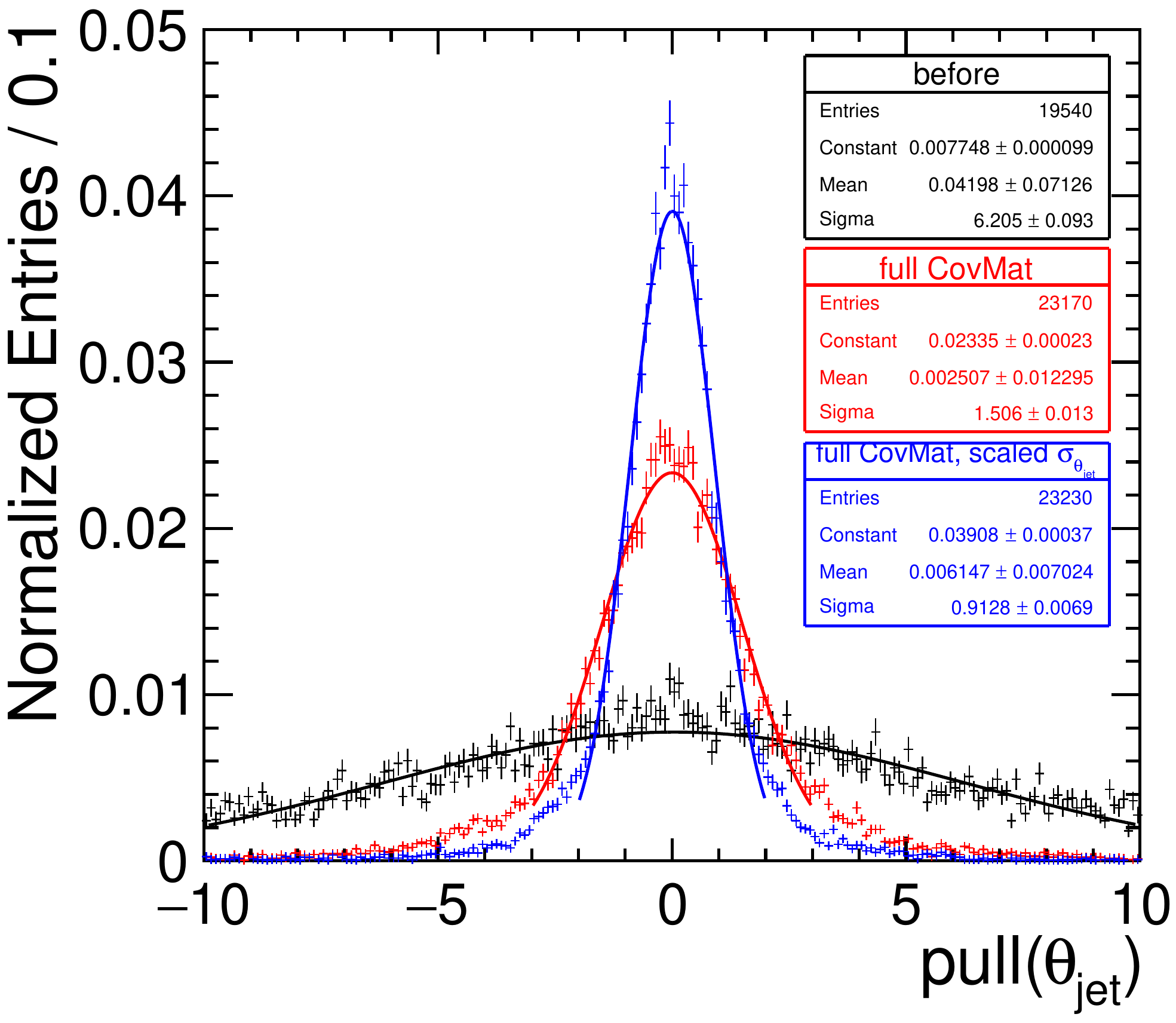}
		\caption{}
		\label{fig:Pull_jetTheta}
	\end{subfigure}
	\caption{(a) Normalized residual of $\theta_{\mathrm{PFO}}$ for charged particles (black), photons (blue) and neutral hadrons (red) (b) Pull of the jet angle ($\theta_{\mathrm{jet}}$) from the kinematic fit when the jet angular uncertainty ($\sigma_{\theta_{\mathrm{jet}}}$) receives contribution from only charged particles (black), all particles (red) and when $\sigma_{\theta_{\mathrm{PFO}}}$ is scaled up for photons and neutral hadrons by a factor of 1.3 and 1.8, respectively, motivated from fig.~\ref{fig:sigmaTheta_pfo}}
	\label{fig:sigmaTheta}
\end{figure}

It was shown that by tagging charged leptons in $b$- or $c$-jets, the semi-leptonic decays can be found \cite{radkhorrami2021conceptual}. The momentum of the missing neutrino can then be obtained up to a sign ambiguity, using the kinematics of the semi-leptonic decay including the four-momentum of visible decay products of the semi-leptonic decay, the mass, and the flight direction of the parent hadron. Imposing constraints on energy and momentum conservation and/or invariant masses of known particles will resolve the sign ambiguity by a kinematic fit. Full details of the measurement uncertainties for each particle flow object (PFO) as reconstructed particle candidate are provided by the Pandora particle flow algorithm \cite{Thomson_2009} used for event reconstruction in the ILD concept. The energy and spatial resolutions of each sub-detector of ILD are propagated to full covariance matrix for each PFO. While angular errors of the charged particles are very well modeled using the ILD tracker (black histogram in fig.~\ref{fig:sigmaTheta_pfo}), the angular uncertainties for the photons and neutral hadrons (blue and red histograms in fig.~\ref{fig:sigmaTheta_pfo}) should be scaled up by a factor of 1.3 and 1.8, respectively, when they are propagated to form the initial covariance matrix of the jet ($\sigma_{\mathrm{det}}$).
The jet energy error has been studied in \cite{radkhorrami2021conceptual,radkhorrami2021kinematic,Ebrahimi:2017}. In the previous jet error parameterization \cite{Ebrahimi:2017}, only charged particles contributed to the jet angular uncertainties (black histogram in fig.~\ref{fig:sigmaTheta}b) and the kinematic fit hypothesis was too tight for adjusting the four-momentum of jets. Therefore, a fixed angular uncertainty ($\sigma_{\theta_{\mathrm{jet}}}$=100 mrad) was assumed for all jets \cite{Ebrahimi:2017}. In the new jet error parameterization, all particles contribute to the jet angular uncertainties and hence, the angular uncertainties are parameterized for individual jets. This significantly improves the pull distribution of the jet angles (red histogram in fig.~\ref{fig:Pull_jetTheta}) and the gain is even more drastic when $\sigma_{\theta_{\mathrm{PFO}}}$ is scaled-up for photons and neutral hadrons (blue histogram in fig.~\ref{fig:Pull_jetTheta}). In analogous to $\theta_{\mathrm{jet}}$, the azimuthal angle of the jet ($\phi_{\mathrm{jet}}$) has similar behavior for photons and neutral hadrons.
\section{$Z/H$ mass reconstruction}\label{SEC:ZHMassReco}
\begin{figure}[htbp] \centering
	\includegraphics[width=0.7\textwidth]{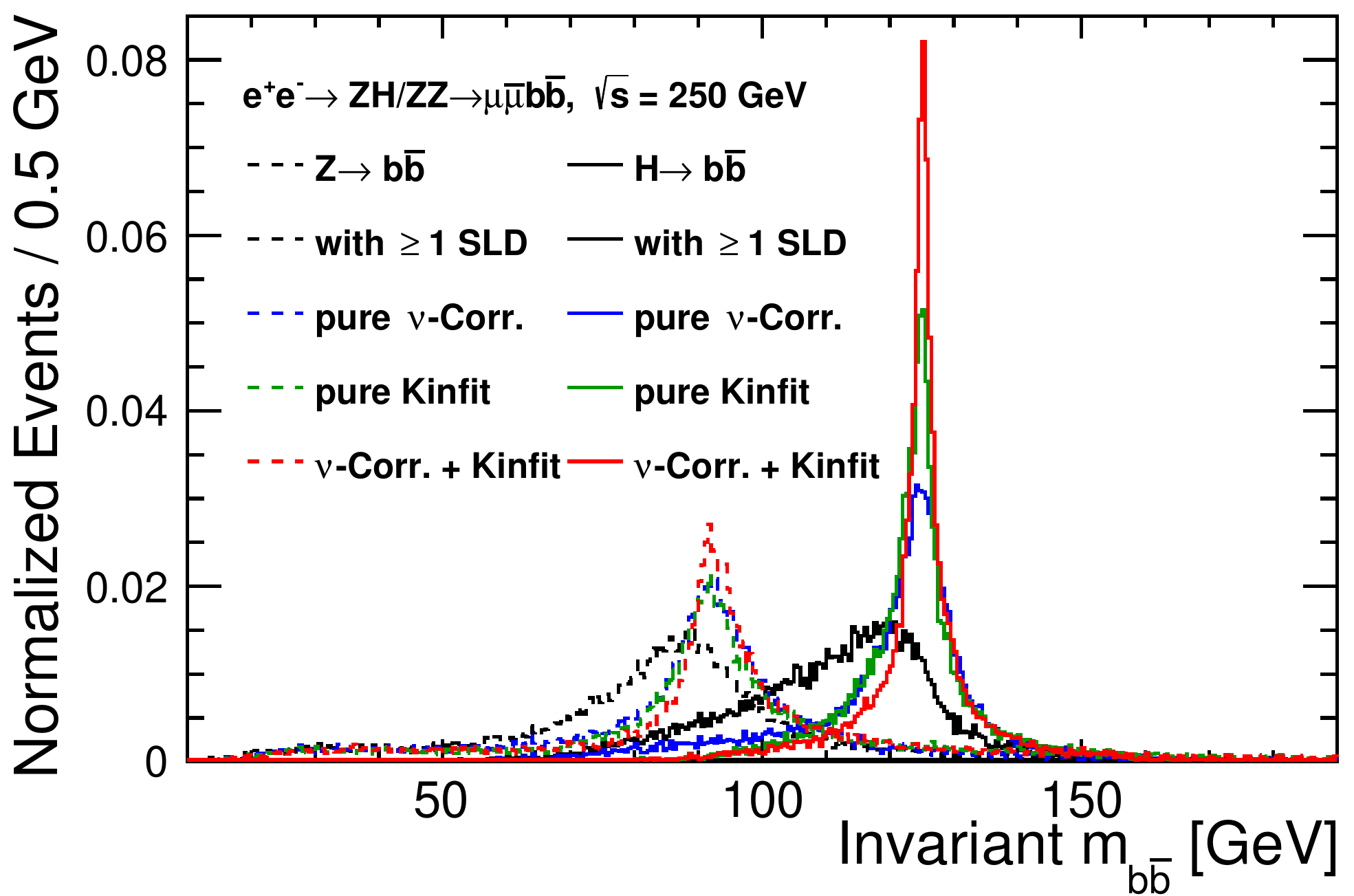}
	\caption{Invariant di-jet mass in $ZH$ and $ZZ$ events reconstructed by the PFA only (black), with neutrino correction (blue), with kinematic fit (green) and with neutrino correction and kinematic fit (red).}
	\label{fig:ZZHZ_InvMass}
\end{figure}

The neutrino correction -- for the time being, based on cheated inputs in terms of decay point and mass of the mother hadron, and association and momenta of the visible decay products -- in addition to the new jet error parameterization, is applied on the introduced $ZH/ZZ\rightarrow\mu\bar{\mu}b\bar{b}$ samples. The mass recovery by the neutrino correction alone is shown in the blue histograms in fig.~\ref{fig:ZZHZ_InvMass} for both Higgs boson (solid) and $Z$ boson (dashed). Here the sign ambiguity is resolved by the kinematic fit, but the di-jet invariant mass is obtained from the pre-fit jet four-momenta. Even without applying the neutrino correction, the di-jet invariant mass can be significantly improved by the new jet error parametrization described in section \ref{SEC:KinFit} fed to the kinematic fit (the green histograms in fig.~\ref{fig:ZZHZ_InvMass}). Finally, the combination of the neutrino correction and the kinematic fit is shown in the red histograms in fig.~\ref{fig:ZZHZ_InvMass}. The performance of the combination of the kinematic fit and the neutrino correction based
on reconstructed information is expected somewhere between the green and red histograms even if the neutrino correction is less powerful when using reconstructed information only. In all cases, the problem of pulling the $Z$ peak towards the Higgs peak shown in fig.~\ref{fig:ZZHZ_Higgsmass} is avoided and the $ZH$/$ZZ$ separation is significantly improved with respect to the black histograms. This will benefit any Higgs studies where event selection is done based on di-jet masses. The impact of the neutrino correction seems much smaller for $Z\rightarrow b\bar{b}$ than for $H\rightarrow b\bar{b}$. This is suspected to be due to the natural width of the $Z$ boson, which is similar in size to the effects of detector resolution.
\section{Conclusions and Outlook}\label{SEC:Concl}
The invariant di-jet mass plays an important role in the separation of the $ZZH$ background and the $ZHH$ signal for the Higgs self-coupling analysis. Since $H\rightarrow b\bar{b}$ is the predominant decay mode of the Higgs boson, the reconstruction of heavy-flavor jets highly affects the invariant di-jet mass. A combination of a kinematic fit with a conceptual approach for a correction of semi-leptonic decays based on a detailed reconstruction of the decay kinematics can improve the heavy flavour jet reconstruction. The propagation of PFO-level measurement uncertainties for all reconstructed particles obtains a precise knowledge of the measurement errors for each individual jet.  The neutrino correction, for the time being, based on cheated inputs as proof-of-principle, and the kinematic fit with detailed error parameterization have been applied on $ZH/ZZ\rightarrow\mu\bar{\mu}b\bar{b}$ events. Each of the techniques considerably enhances the $ZH$/$ZZ$ separation via the di-jet mass and their combination shows significant improvement. The neutrino correction based on fully reconstructed information is ongoing and both techniques will be applied to the Higgs self-coupling analysis in the future.
\printbibliography[title=References]

@misc{theildcollaboration2020international,
	author		= {The ILD Collaboration},
	title		= {International Large Detector: Interim Design Report}, 
	year		= {2020},
	eprint		= {2003.01116},
	archivePrefix	= {arXiv},
	primaryClass	= {physics.ins-det}
}

@article{List:88030,
	author		= {Benno List and Jenny List},
	title		= {{M}arlin{K}infit: {A}n {O}bject--{O}riented {K}inematic {F}itting {P}ackage},
	journal	= {LC Notes},
	address	= {Hamburg},
	publisher	= {DESY},
	reportid	= {PHPPUBDB-10294, LC-TOOL-2009-001},
	year		= {2009},
	cin		= {FLC},
	cid		= {I:(DE-H253)FLC-20120731},
	pnm		= {Facility (machine) ILC $R\&D$ (POF1-ILC-20130405)},
	pid		= {G:(DE-H253)POF1-ILC-20130405},
	experiment	= {EXP:(DE-H253)ILC(machine)-20150101},
	typ		= {PUB:(DE-HGF)25},
	url		= {https://bib-pubdb1.desy.de/record/88030}
}

@phdthesis{Duerig:310520,
	author		= {D{\"u}rig, Claude Fabienne},
	title		= {Measuring the Higgs Self-coupling at the International Linear Collider},
	issn		= {1435-8085},
	school		= {Uni. Hamburg},
	type		= {Dr.},
	address	= {Hamburg},
	othercontributors = {List, Jenny and Garutti, Erika},
	publisher	= {Verlag Deutsches Elektronen-Synchrotron},
	reportid	= {PUBDB-2016-04283, DESY-THESIS-2016-027},
	series		= {DESY-THESIS},
	pages		= {246},
	year		= {2016},
	note		= {Universit{\"a}t Hamburg, Diss., 2016},
	cin		= {FLC},
	cid		= {I:(DE-H253)FLC-20120731},
	pnm		= {611 - Fundamental Particles and Forces (POF3-611) / SFB 676-B01 - Optimising the ILC setup: Physics programme, running scenarios and design choices (DFG-SFB-676-B01)},
	pid		= {G:(DE-HGF)POF3-611 / G:(DE-H253)DFG-SFB-676-B01},
	experiment	= {EXP:(DE-MLZ)NOSPEC-20140101},
	typ		= {PUB:(DE-HGF)3 / PUB:(DE-HGF)29 / PUB:(DE-HGF)11},
	doi		= {10.3204/PUBDB-2016-04283},
	url		= {https://bib-pubdb1.desy.de/record/310520}
}

@article{Thomson_2009,
	author		= {Thomson, M.A.},
	title		= {Particle flow calorimetry and the PandoraPFA algorithm},
	volume		= {611},
	ISSN		= {0168-9002},
	url		= {http://dx.doi.org/10.1016/j.nima.2009.09.009},
	DOI		= {10.1016/j.nima.2009.09.009},
	number		= {1},
	journal	= {Nuclear Instruments and Methods in Physics Research Section A: Accelerators, Spectrometers, Detectors and Associated Equipment},
	publisher	= {Elsevier BV},
	year		= {2009},
	month		= {Nov},
%	pages		= {25–40}
}

@phdthesis{Ebrahimi:2017,
	author		= {Aliakbar Ebrahimi},
	title		= {Jet Energy Measurements at ILC: Calorimeter DAQ Requirements and Application in Higgs Boson Mass Measurements},
	reportNumber	= {PUBDB-2017-11891},
	school		= {DESY},
	year		= {2017},
	doi		= {10.3204/PUBDB-2017-11891},
	url		= {https://bib-pubdb1.desy.de/record/394104}
}

@misc{radkhorrami2021kinematic,
	title		= {Kinematic Fitting for ParticleFlow Detectors at Future Higgs Factories}, 
	author		= {Yasser Radkhorrami and Jenny List},
	year		= {2021},
	eprint		= {2110.13731},
	archivePrefix	= {arXiv},
	primaryClass	= {hep-ex}
}

@misc{radkhorrami2021conceptual,
	author		= {Yasser Radkhorrami and Jenny List},
	title		= {Conceptual aspects for the improvement of the reconstruction of $b$- and $c$-jets at $e^{+}e^{-}$ Higgs Factories with ParticleFlow detectors}, 
	year		= {2021},
	eprint		= {2105.08480},
	archivePrefix	= {arXiv},
	primaryClass	= {hep-ex}
}
\end{document}